\documentclass[twocolumn]{aastex631}

\usepackage{siunitx}
\usepackage{bm}
\usepackage{mathtools}
\usepackage{CJK}         % Chinese name

\begin{document}
\begin{CJK*}{UTF8}{gbsn}

\title{Simulating X-ray Reverberation in the UV-Emitting Regions of Active Galactic Nuclei Accretion Disks with 3D Multi-Frequency Magnetohydrodynamic Simulations}

\author[0000-0002-1174-2873]{Amy Secunda}
\affil{Department of Astrophysical Sciences, Princeton University, Peyton Hall, Princeton, NJ 08544, USA}

\author[0000-0002-2624-3399]{Yan-Fei Jiang (姜燕飞)}
\affil{Center for Computational Astrophysics, Flatiron Institute, New York, NY 10010, USA}

\author[0000-0002-5612-3427]{Jenny E. Greene}
\affil{Department of Astrophysical Sciences, Princeton University, Peyton Hall, Princeton, NJ 08544, USA}

\begin{abstract}
    Active galactic nuclei (AGN) light curves observed with different wavebands show that the variability in longer wavelength bands lags the variability in shorter wavelength bands. Measuring these lags, or reverberation mapping, is used to measure the radial temperature profile and extent of AGN disks, typically with a reprocessing model that assumes X-rays are the main driver of the variability in other wavelength bands. To demonstrate how this reprocessing works with realistic accretion disk structures, we use 3D local shearing box multi-frequency radiation magnetohydrodynamic (MHD) simulations to model the UV-emitting region of an AGN disk, which is unstable to the magnetorotational instability (MRI) and convection. At the same time, we inject hard X-rays ($>1$~keV) into the simulation box to study the effects of X-ray irradiation on the local properties of the turbulence and the resulting variability of the emitted UV light curve. We find that disk turbulence is sufficient to drive intrinsic variability in emitted UV light curves and that a damped random walk (DRW) model is a good fit to this UV light curve for timescales $>5$~days. Meanwhile, the injected X-rays have almost no impact on the power spectrum of the emitted UV light curve. In addition, the injected X-ray and emitted UV light curves are only correlated if there is X-ray variability on timescales $>1$~day, in which case we find a correlation coefficient $r=0.52$. These results suggest that hard X-rays with scattering dominated opacity are likely not the main driver of the reverberation signals.

\end{abstract}

\section{Introduction}
Although active galactic nuclei (AGN) are among the most luminous objects in the Universe, due to their great distance it is generally not possible to resolve the microarcsecond scales of their accretion disks, broad line regions (BLRs), or event horizons. However, because AGN light curves vary over a wide range of frequencies, temporal resolution can be used in place of spatial resolution. A primary technique used to study AGN, called reverberation mapping, makes use of this temporal resolution \citep{Peterson:2014}.

Reverberation mapping was first proposed by \cite{Blandford1982} as a way to measure distances to the BLR by finding a time lag between variability in continuum light curves and variability in broad emission lines \citep[see also][]{Kaspi_2000,Peterson2004,Bentz_2015,Grier:2017}. Disk continuum reverberation mapping can similarly be used to measure the radial extent of the accretion disk itself, because X-rays emitted by the corona move outwards along the disk and are absorbed and re-emitted by different temperature regions of the disk at different wavelengths based on the local temperature \citep[e.g.,][]{Sergeev:2005,Cackett:2007,Cackett:2018,Cackett:2021,derosa2015,Edelson2015,Edelson2017,Edelson2019,Jiang:2017,Fausnaugh2016,Starkey2017,Homayouni:2022}. Therefore, there will be a lag on the light-crossing timescale between variability in light curves from short wavelength bands and light curves that have been reprocessed and re-emitted in longer wavelength bands. This lag is used to determine the distance between different temperature regions of the accretion disk.

Traditional models for reverberation mapping make the assumption that the variability in UV and optical AGN light curves is driven by the variability of the X-ray irradiation. However, there is significant evidence for additional variability in UV-optical light curves that does not appear to come from X-ray light curves \citep[e.g.,][]{Uttley:2003,Arevalo:2009,Edelson:2019,Yu:2022,Hagen2023,Cackett:2023,Kara:2023}. Instead this variability could come from fluctuations in the UV-optical regions of the accretion disk due to magnetorotational instability (MRI) driven turbulence \citep{BalbusHawley1991} or convection resulting from the enhanced opacity in the UV-optical region of the disk \citep{JiangBlaes2020}. 

While there has been considerable observational work done using reverberation mapping, theoretical work to model the underlying physics of the reprocessing of hard irradiation into the softer radiation emitted by the disk has lagged behind. Most of the previous work modeling this reprocessing is (semi-)analytic and makes assumptions about the vertical profiles of various disk parameters, which often remain fixed over time or are not-calculated self-consistently to include the interaction of radiation and gas \citep[e.g.,][]{Sun:2020,Kammoun:2021b,Salvesen:2022,Panagiotou:2022}. The recent development of accurate, high-speed, numerical methods to calculate radiative transfer coupled to magnetohydrodynamics (MHD) make it possible for the first time to examine how both radiation and magnetic fields have an impact on AGN disk structure and turbulence \citep{Jiang2013,JiangBlaes2020,Jiang2021}. Adding realistic Rosseland and Planck mean opacities that include line opacities instead of only electron scattering and free-free absorption also has an impact on driving convection in AGN disk models \citep{JiangBlaes2020}. However, all of these MHD simulations model radiation using a single frequency group, with the radiation integrated over all frequencies, and do not examine the impact of X-ray irradiation on the disk and the light curves emitted by the disk.

In this paper we provide an early application of a multi-frequency radiation MHD code, which instead of integrating radiation over all frequencies, integrates radiation over multiple frequency groups \citep{Jiang2022}. This multi-frequency capability allows us to simulate from first principles the reprocessing of high-frequency light incident on an accretion disk into light emitted in the UV based on the local disk temperature to better understand the drivers of variability in UV-optical AGN light curves. By first principles, we mean that the radiation energy and momentum are coupled to the gas evolution and that the temperature, opacity, and energy of the simulation depend self-consistently on the local gas density and pressure. In Section \ref{sec:methods} we describe the methods we use for our simulations. In Section \ref{sec:results} we present, compare, and discuss the simulated disks and light curves from our simulations with and without injected X-rays. Finally, in Section \ref{sec:discuss} we summarize our results and present ideas for future work.

\section{Methods}
\label{sec:methods}
Using {\sc athena++} we solve the ideal MHD equations coupled with time-dependent, implicit radiative transfer equations for intensities over discrete angles in the manner described in \cite{Jiang2022}, which builds on \cite{Jiang2021} by allowing for multiple frequency groups. We use two frequency groups, one group covering the frequency range $[0,1\text{kev}]$ to model the radiation field emitted by the disk locally (at UV wavelengths) and the other group covering the frequency range $(1\text{keV},\infty)$ for the X-ray irradiation. Notice that we turn off the frequency shift due to Doppler effect as its effect is small and can cause numerical diffusion in the frequency space for the two groups we adopt.  

We model a local patch of the accretion disk around a supermassive black hole with mass $M_{\rm BH}=5\times 10^7~M_{\odot}$ under the local shearing box approximation with the box centered at $r_0=45R_{\rm s}$, where  $R_{\rm s}$ is the Schwarzschild radius. The corresponding Keplerian orbital frequency, $\Omega=4.73\times10^{-6}$~s$^{-1}$ and shear parameter, $q = - d\ln \Omega /d\ln r = 3/2 $. We choose our simulation domain to be $(x,y,z)\in (-1.5H_g,1.5H_g)\times(-6H_g,6H_g)\times (-48H_g,48H_g)$ with resolution $N_x=48$, $N_y=192$ and $N_z=1536$. Here $H_{\rm g}=\num{1.11e12}$~cm is the gas pressure scale height of the disk, while the total pressure scale height (including radiation pressure) will be $H_{\rm r}=6.89 H_{\rm g} = \num{7.65e12}$~cm. We use the periodic and shear periodic {\sc athena++} boundary conditions for the $x$- and $y$-boundaries, respectively. For the $z$-boundaries, we use outflow boundary conditions for the gas. Intensities for the low frequency group are allowed to leave the simulation domain but they are not allowed to enter. For the high frequency group, we set the incoming intensities based on a prescribed X-ray light curve. Outgoing intensities are copied from the last active zone to the ghost zone. 

We first perform a one-frequency run to model only the locally emitted UV photons without X-ray irradiation. Then we restart the simulation with two frequency groups to model the locally generated UV photons and X-ray irradiation simultaneously. We use $N=48$ discrete angles in each cell for each frequency group to resolve the angular distribution of the radiation field. Inside the simulation box we initialize the UV radiation isotropically based on the local disk temperature and the X-ray radiation isotropically based on a constant radiation energy, $E_{\rm r,1}=10^{-4}E_{\rm r,0}^{\rm mid}$, where $E_{\rm r,0}^{\rm mid}$ is the midplane radiation energy for the UV frequency group.

We inject two different X-ray light curves for two otherwise initially identical two-frequency runs (Run A and Run B). The time-dependence of the intensity of this radiation is determined by a mock X-ray AGN light curve that we model as a damped random walk (DRW) as in \cite{Kelly2009}, because the DRW model has been shown to be a decent representation of AGN light curves for timescales of days to years \citep[e.g.][]{Kelly2009, Kozlowski2010, MacLeod2012, Zu2013}. For Run A, we use a damping timescale $\tau_{\rm damp}=30$~days and a time-sampling of $dt=1$~day, for Run B we use $\tau_{\rm damp}=0.1$~days and a time-sampling of $dt=0.001$~days. Run A will have more variability at low frequencies than Run B, while at high frequencies Run B will have more variability. 

We found that if we normalized the intensity of the injected X-rays to the time-averaged total UV flux leaving the top and bottom of the simulation box in the one-frequency run, the X-rays do not have a large impact on our simulation, at least over a few thermal timescales. Since we are interested in X-ray reprocessing, we instead normalize the X-ray irradiation to eight times the average UV emission. In the future a parameter study is needed to determine how this normalization impacts the reprocessing. Finally, we average the injected flux over all angles in our simulation and inject this averaged intensity along all angles pointing into the box.

We determine the timestep-dependent opacity for the UV component in each cell with a bi-linear interpolation of the OPAL opacity tables from \cite{IglesiasRogers1996} based on each cell's current gas temperature and density. For our X-ray component, we take the proper average of the free-free absorption opacity and electron scattering opacity, which ends up to be dominated by the scattering value.

We set the initial vertical profile of our simulations using the equations of radiation hydrostatic equilibrium, as in \cite{Jiang2016}. Initial and normalization parameters are given in Table \ref{tab:units}. For our initial conditions the ratio of the radiation to the gas pressure is $P_{\rm rat}=4.4$. We initialize the magnetic field in our simulation as in \cite{Jiang2013} as two oppositely twisted flux tubes with zero net-flux and set the ratio of the gas pressure to the magnetic pressure to $\beta=200$ in the midplane. We also add a purely vertical magnetic field with $\beta = 2 \times 10^4$ to increase the Maxwell stress in order to achieve a temporally averaged $\dot M/\dot M_{\rm edd} \approx 2 \times 10^{-2}$ once the simulation reaches steady state.

\begin{table}[]
    \centering
    \begin{tabular}{c|c}
        Parameter & Unit \\
        \hline
        \bm{$\Omega_0$} & $\num{4.73e-6}$~s$^{-1}$\\
        \bm{$\rho_0$} & $10^{-7}$~g~cm$^{-3}$ \\
        \bm{$T_0$} & \num{2e5}~K \\
        \bm{$H_{\rm g}$} & $\num{1.11e12}$~cm \\
        $H_{\rm r}$ & 6.89~$H_{\rm g}=\num{7.65e12}$~cm \\
        $F_{\rm max}$ & \num{5.80e12}~erg~s$^{-1}$~cm$^{-2}$ \\
        $\tau_0$ & $\num{6.26e4}$ \\
        $P_{\rm g,0}$ & \num{2.77e6}~dyn~cm$^{-2}$ \\
        $P_{\rm rat}=P_{\rm r,0}/P_{\rm g,0}$  & 4.37 \\
        $c_{\rm rat}=c/c_{\rm s}$  & \num{5.69e3}\\
        $r_0$ & $r_0=45$~R$_{\rm s}=\num{6.68e14}$~cm \\
    \end{tabular}
    \caption{$\Omega_0$, $\rho_0$, $T_0$, and $H_{\rm g}$ are fiducial units we use as normalization parameters in our simulation, for time, mass, temperature, and distance, respectively. $H_{\rm r}$ is a fiducial unit for the total (gas and radiation combined) pressure scale height. $F_{\rm max}$, $\tau_0$, and $P_{\rm g,0}$, are the initial $F_{r,z}(\tau<1)$, midplane optical depth, and gas pressure, respectively. $P_{\rm rat}$ and $c_{\rm rat}$ are the initial ratio of the radiation pressure to the gas pressure and the ratio of the speed of light to the sound speed, respectively, at the midplane. $r_0$ is the fixed location of the center of the shearing box.}
    \label{tab:units}
\end{table}

\begin{figure*}
    \centering
    \includegraphics[width=\textwidth]{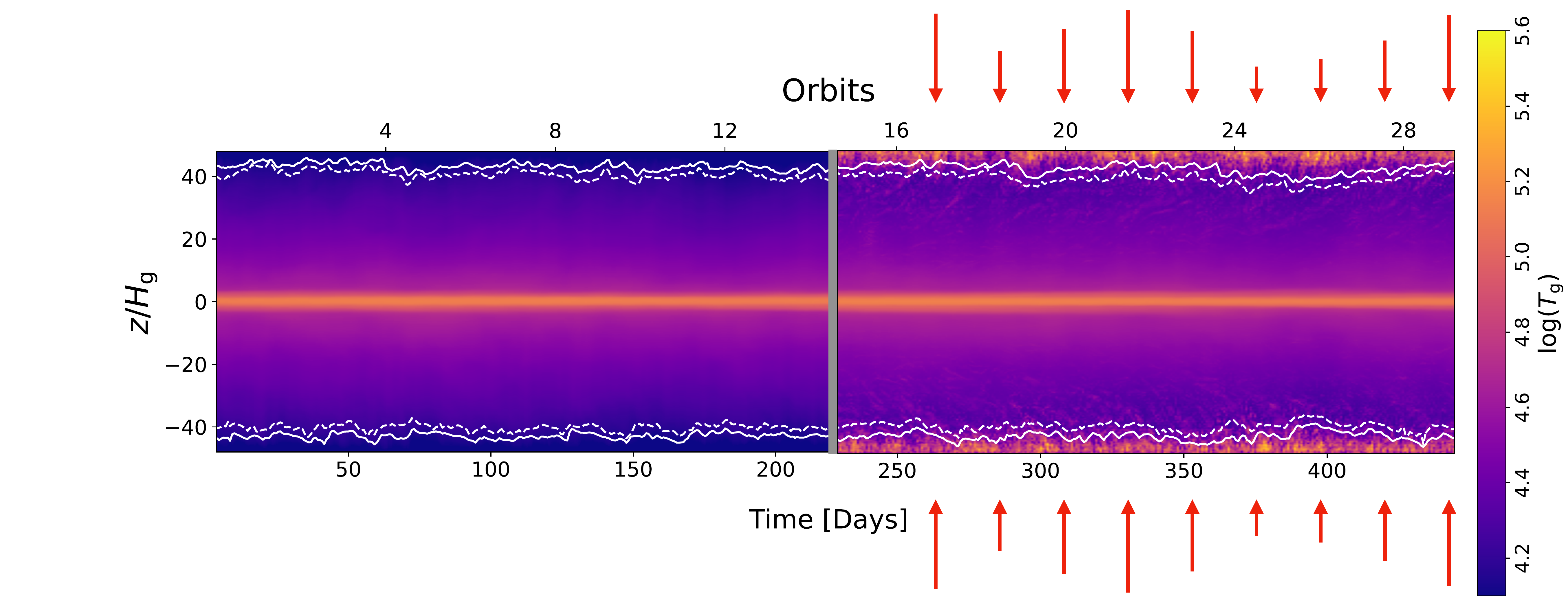} \\
    \includegraphics[width=0.94\textwidth]{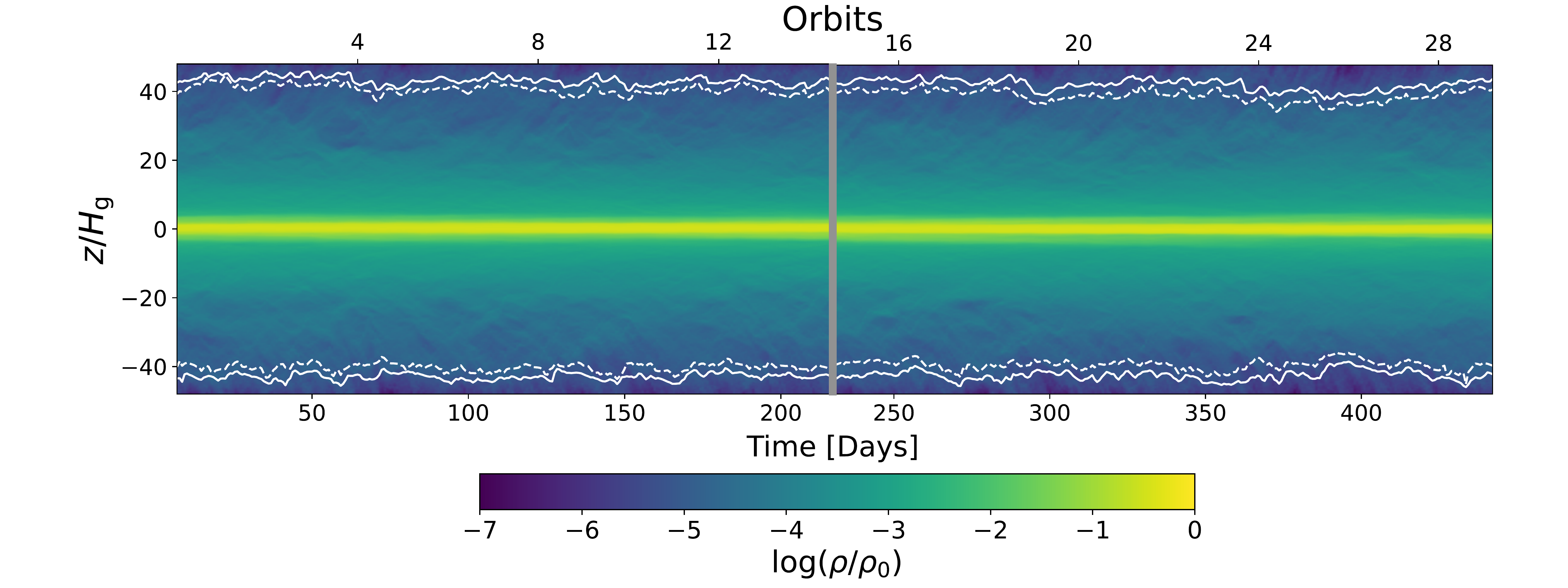} \\
    \includegraphics[width=0.94\textwidth]{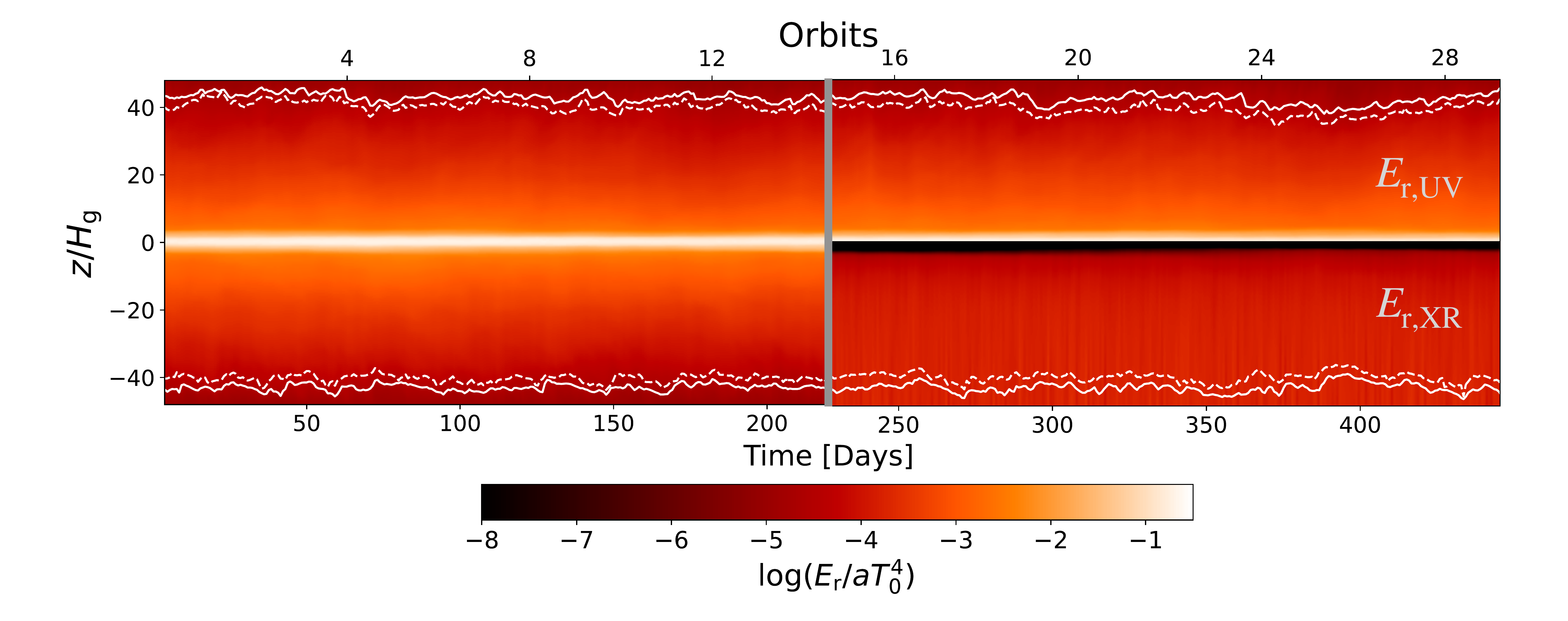}
    \caption{\textbf{Top panel:} Schematic of our simulation and space-time diagram of the gas temperature in the one- and two-frequency simulations in the region left and right of the grey line, respectively. The red arrows represent the time-varying X-ray radiation we inject in through the upper and lower boundaries. \textbf{Middle panel:} Space-time diagram of the gas density in the one- and two-frequency simulations in the region left and right of the grey line, respectively.
    \textbf{Bottom panel:} Same as middle panel, but for the radiation energy. We show the UV radiation energy above the midplane and the X-ray radiation energy below the midplane. In all panels the solid white line is $z(\tau_{\rm UV}=1)\approx z(\tau_{\rm XR}=1)$ and the dashed white line is $z(\tau_{a,\rm UV}=1)$. $\tau_{a,\rm XR}<1$ throughout our simulation.}
    \label{fig:time_avg}
\end{figure*}

\section{Results}
\label{sec:results}

\subsection{Disk Structure}
\label{sec:disk_structure}

We show the horizontally-averaged vertical profiles as a function of time for the gas temperature, gas density, and radiation energy for the one-frequency run (to the left of the grey line) and two-frequency run B (to the right of the grey line) in Figure \ref{fig:time_avg}. In the top panel the red arrows represent the time varying X-ray irradiation we inject into the top and bottom of our simulation box in our two-frequency runs. The solid white lines show the location of the UV photosphere, which is very similar to the location of the X-ray photosphere, because both optical depths are dominated by the scattering opacity near the surface of the disk ($z(\tau_{\rm UV}=1) \approx z(\tau_{\rm XR}=1)$). Absorption optical depth for the X-ray component is smaller than 1 throughout the whole disk, but the effective absorption optical depth is $\sim 1.3$. The vertical profiles for two-frequency run A, not shown here, are qualitatively the same as in run B.

The vertical profiles are very similar between the one- and two-frequency runs, except for the profile for the gas temperature (top panel), which is higher throughout the simulation when we inject X-rays. Unsurprisingly, the temperature increases the most in the two-frequency run in the photosphere, where on short timescales the temperature can exceed the midplane temperature, and inwards an additional $10~H_{\rm g}$, where the temperature can nearly reach $T_{\rm g}\sim 10^5$~K. In contrast to isothermal simulations, we do no see any sign of disk winds in our simulations \citep[e.g.,][]{Bai:2013}.

In the bottom panel of Figure \ref{fig:time_avg}, we show the radiation energy for the low- and high-frequency group above and below the midplane, respectively, to the right of the grey line. The midplane radiation energy is dominated by the energy from the low-frequency group. However, at around $\pm30~H_{\rm g}$, the radiation energy from the high-frequency group becomes the dominant radiation energy. On average, the energy of the X-ray radiation manages to penetrate down to $\tau_{\rm XR} \approx 240$ or $\pm 6.5$~$H_{\rm g}$ before it drops by an e-folding.

Overall, the $\alpha$-viscosity parameter in our simulations ranges from around 0.025 to 0.04. In all runs, the Maxwell stress is about a factor of 6 larger than the Reynolds stress. The average thermal timescale is $\tau_{\rm therm} = 1/(\alpha \Omega) \approx 80$ in all runs.

\subsection{Intrinsic Variability in UV Light Curves}
\label{sec:uv}

\begin{figure*}
    \includegraphics[width=0.49\textwidth]{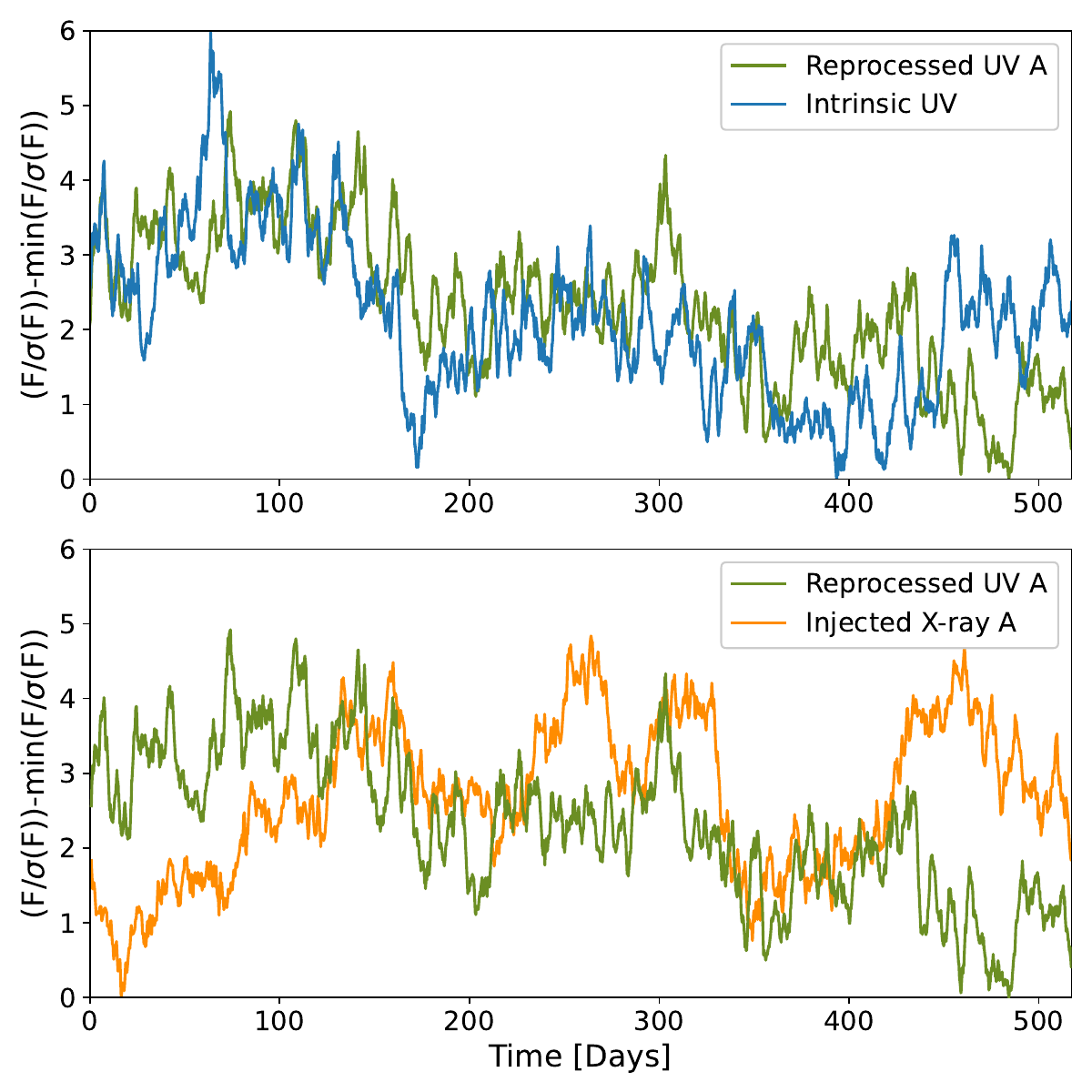}
    \includegraphics[width=0.49\textwidth]{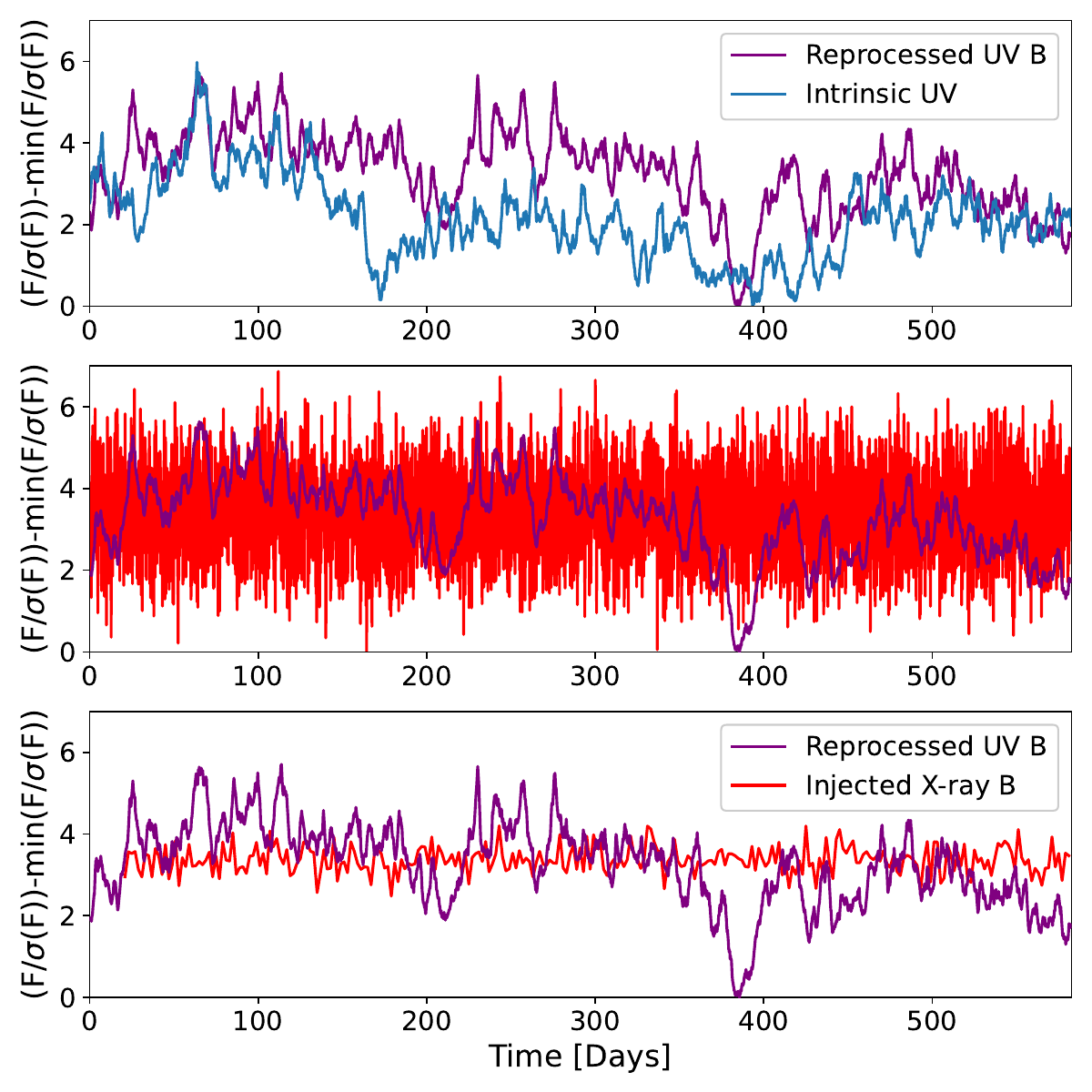} 
    \caption{Top panels compare the UV light curves from the one- and two-frequency runs. Middle and bottom panels compare the UV light curves and injected X-rays from the two-frequency runs. $\tau_{\rm damp}=30$~days for the injected X-ray light curve in the left panels and $\tau_{\rm damp}=0.1$~days for the injected X-ray light curve in the right panels. In the bottom right panel the X-ray light curve has been window-averaged over 2 days for easier comparison. To make each UV light curve, we sum the total outgoing flux from the low-frequency group over every cell at the top and bottom of the simulation box. For easier comparison, we scale the flux, $F$, by the standard deviation of the flux, $\sigma(F)$, and then subtract off the minimum flux for each light curve. Time zero corresponds to when we start injecting the X-rays.}
    \label{fig:lcs}
\end{figure*}

\begin{figure*}[t!]
    \centering
    \includegraphics[width=0.49\textwidth]{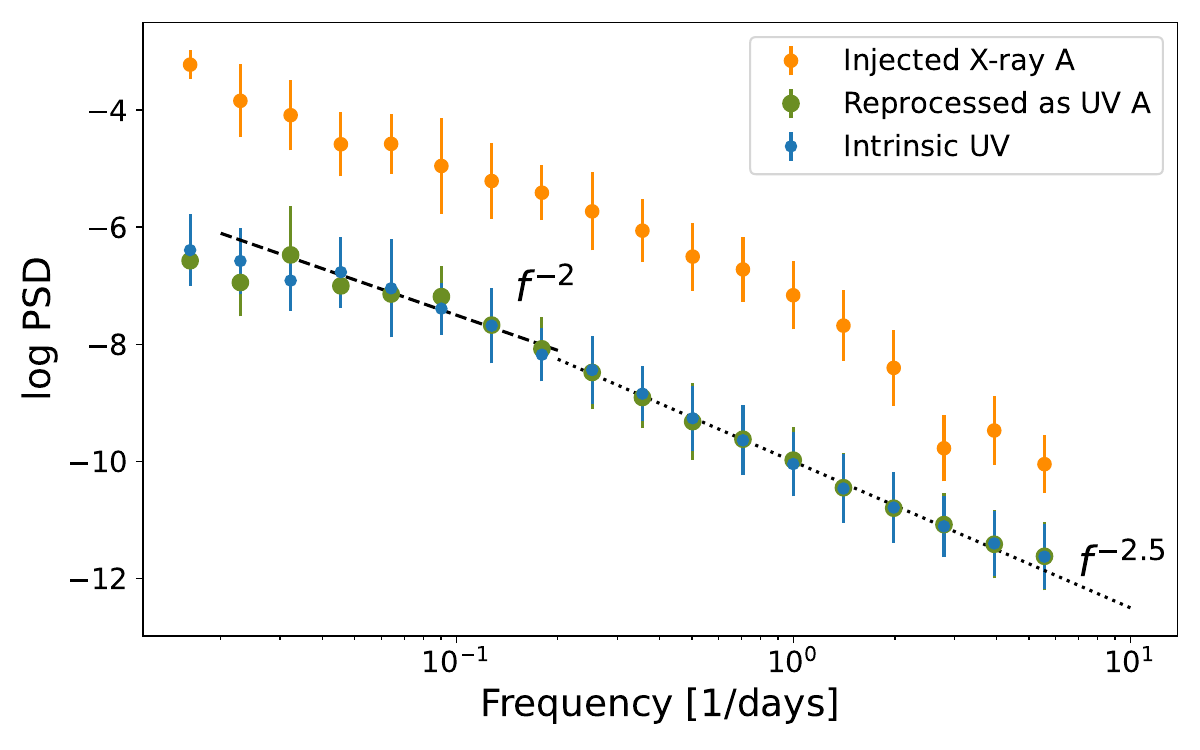}
    \includegraphics[width=0.49\textwidth]{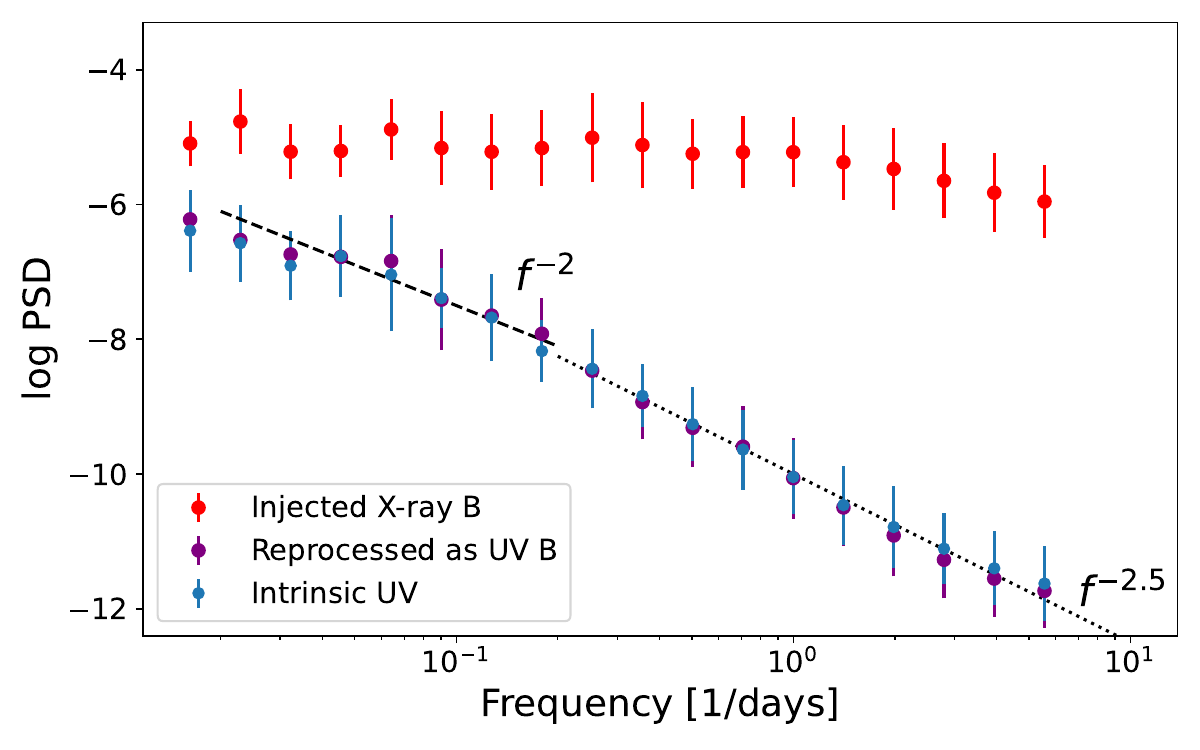} \\
    \includegraphics[width=0.49\textwidth]{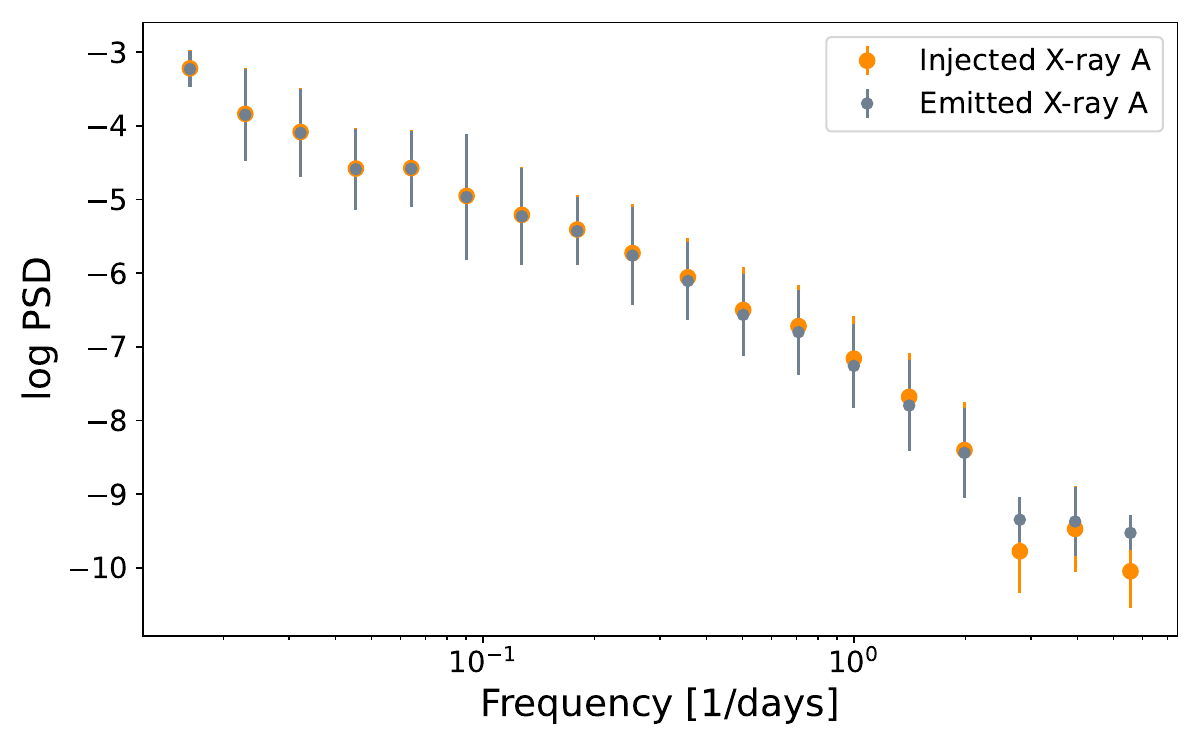}
    \includegraphics[width=0.49\textwidth]{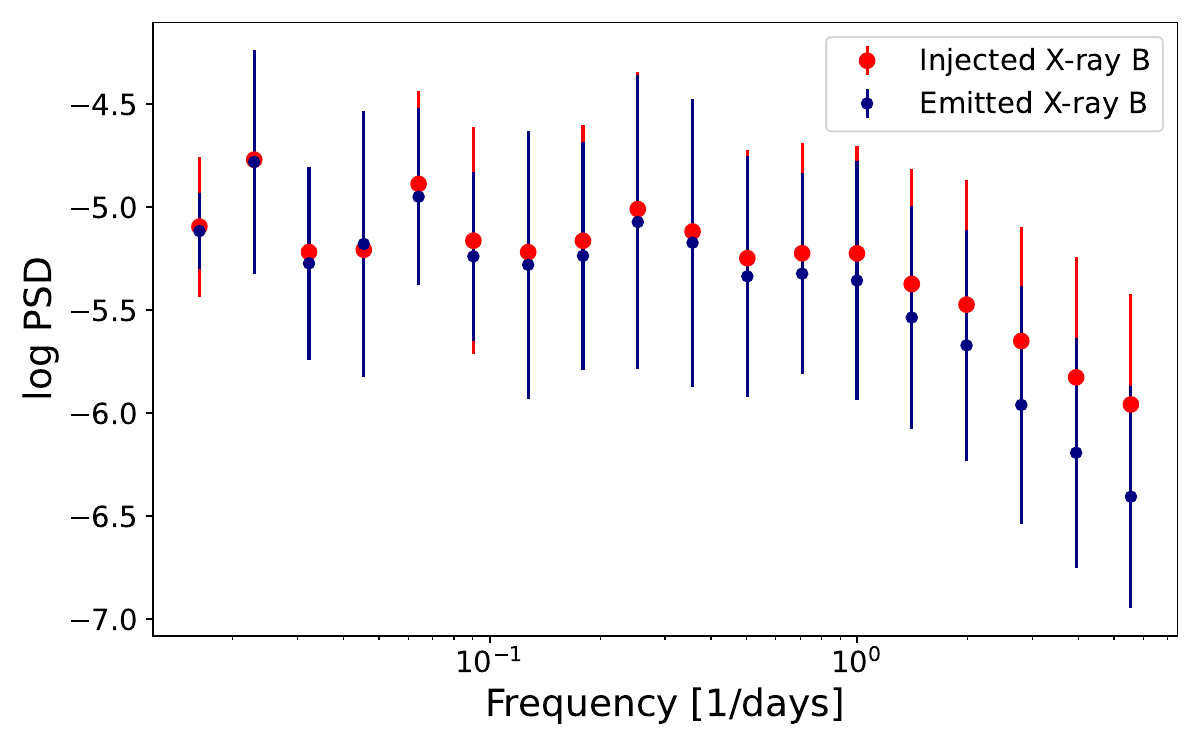}
    \caption{The top panels show the PSDs of the one-frequency UV, the two-frequency UV, and injected X-ray light curves in Figure \ref{fig:lcs}. The bottom panels compare the PSDs of the injected and emitted X-rays for both two-frequency runs. The left (right) panels show the PSDs from run A (run B), where $\tau_{\rm damp}=30$~days ($\tau_{\rm damp}=0.1$~days) for the injected X-ray light curve. We see some evidence that the UV PSDs are damped at $f\lesssim 0.03$~days$^{-1}$. At $0.03< f < 0.2$~days$^{-1}$ the slope of the UV PSDs appear to be well fit by $f^{-2}$, shown as the dashed line, which is the DRW slope above the damping frequency. At $f>0.2$~days$^{-1}$  the UV PSDs are well fit by $f^{-2.5}$, shown at the dotted line.}
    \label{fig:powerlaws}
\end{figure*}

We show the intrinsic UV light curve from the one-frequency run without X-ray irradiation and the power spectral density (PSD) of this light curve in blue in the top panels of Figures \ref{fig:lcs} and \ref{fig:powerlaws}, respectively. The intrinsic UV light curve clearly has variability over timescales of monthly to sub-daily that is driven by MRI turbulence and convection. There is evidence that the UV PSD is damped at the lowest frequencies ($f\lesssim1/(30~\rm{days})$), and then goes as $f^{-2}$ for $0.03< f < 0.2$~days$^{-1}$, suggesting that a DRW model might be a good fit for the intrinsic UV light curve at these frequencies. Recent examinations of long cadence optical AGN light curves have suggested that the DRW damping timescale is related to the thermal timescale of the disk \citep{Kelly:2009,Burke2021,Stone:2022}. The damping timescale of around 30~days that we see evidence of in Figure \ref{fig:powerlaws}, is very similar to the thermal timescale near the surface of the disk in our simulations ($\sim 30$~days), but longer light curves are needed to robustly model this damping timescale.

While the DRW model may be a good fit at lower frequencies, the slope of the UV light curve's PSD is steeper, $\sim f^{-2.5}$, than the DRW model at $f\gtrsim1$/(5 days). In recent high cadence observations of AGN light curves from Kepler, on timescales around ten days there is evidence that the PSD slope is steeper than the $f^{-2}$ power-law of the DRW \citep[e.g.,][]{Smith2018}. In our simulations, the break may be at a higher frequency because we are simulating the UV region as opposed to the optical region being probed by Kepler. \cite{Tachibana:2020} proposed that this break comes from the fact that variability will be averaged over different regions of the disk. Our simulations are highly localized and still have a steeper slope at high frequencies solely because disk fluctuations are unable to produce much variability on timescales this short. However, the lack of averaging in our local simulations may be another reason why we do not see a break at frequencies as low as in Kepler light curves. Nonetheless, it is clear that our intrinsic UV light curve has variability on timescales of monthly to sub-daily, due to MRI- and convection-driven turbulence.

\subsection{Hard X-rays Have Little Impact on UV Emission}

We compare the X-ray and UV light curves from the two-frequency runs with X-ray irradiation in the bottom panels of Figure \ref{fig:lcs} and the PSDs of these light curves in the top panels of Figure \ref{fig:powerlaws}. For run A, the Pearson r-coefficient between the injected X-ray light curve with damping timescale $\tau_{\rm damp}=30$~days and the reprocessed UV light curve is $r=0.52$, with a p-value, $p<10^{-5}$. For run B, where the damping timescale of the injected X-ray light curve is  $\tau_{\rm damp}=0.1$~days, we do not get a significant correlation between the X-ray and reprocessed UV light curves, $p=0.54$. This lack of correlation is despite the fact that the injected X-ray flux is a factor of 8 larger than the emitted UV flux, which is larger than typically observed. However, the lack of correlation in run B may be due to the short damping timescale of the X-ray light curve, which makes the X-ray light curve essentially white noise for all timescales of interest and therefore hard to cross-correlate.

For both runs, the injected X-ray light curves appear to have no impact on the PSDs of the reprocessed UV light curves, which are nearly identical to the intrinsic UV light curves from the one-frequency run without X-ray irradiation. This lack of impact is particularly surprising for run B, where at high frequencies the X-ray irradiation has around 6 orders of magnitude more power than the intrinsic UV light curve. The lack of additional high frequency variability in the reprocessed UV light curve in run B is apparent in the light curve itself, as well. 

The reason the UV PSD from run B is not enhanced at high frequencies may be related to the fact that the emitted X-ray light curve in run B has less power at $f\gtrsim1$/(1 day) than the injected X-ray light curve (see bottom right panel of Figure \ref{fig:powerlaws}). The smoothing of high frequency variability in the injected X-ray light curve occurs because the X-ray irradiation gets scattered instead of absorbed by the disk. As a result, the injected and emitted light curves in run B are less strongly correlated, $r=0.89$, than in run A, where $r=0.998$ and there is less high frequency variability to be smoothed. 

Due to the very low absorption opacity for the X-ray irradiation, there is very little absorption and direct heating, as is typically assumed for reverberation mapping. Instead, our results suggest that the scattering of the incident X-ray photons off of electrons leads to a momentum transfer that eventually will thermalize, producing the small amount of disk heating we see in our simulations (see Figure \ref{fig:time_avg}) and the correlation we see between the X-ray and UV light curves in run A. This momentum transfer and disk heating does have some impact on the re-emitted UV light curves in run B, as well, which in the top right panel of Figure \ref{fig:lcs} is clearly not identical to the intrinsic UV light curve. For both two-frequency runs, the reprocessed UV light curves are not strongly correlated with the intrinsic UV light curve from the one-frequency run, with $r=0.46$ and $r=0.54$, for run A and run B, respectively. This picture agrees with \cite{Gardner:2017}, who found that UV emission from NGC 5548 was in good agreement with analytic models assuming a puffed up photosphere with low X-ray absorption opacities. Furthermore AGN disks with scattering atmospheres may help account for the longer than anticipated lag timescales of recent AGN reverberation mapping campaigns \citep{Hall:2018}.

The time-dependent opacities in our simulation are calculated self-consistently from the temperature and density determined by evolving the MHD equations using an initial vertical profile in radiation hydrostatic equilibrium. If our simulation box was located in a different temperature region of the disk or we used different AGN parameters, such as supermassive black hole (SMBH) mass or accretion rate, it is possible that we would have different initial conditions that would lead to an increased absorption or decreased scattering optical depth near the surface of the disk. We have also assumed that the gas is fully ionized, which may not be true if the temperature near the disk surface drops below $T_{\rm g} \approx 10^4$~K, which it might for different initial conditions. A lower ionization fraction would decrease the scattering opacity and likely lead to more absorption \citep{Garcia:2013}. 

\cite{Salvesen:2022} found there was a non-zero thermalization time for the reprocessing of X-ray irradiation. We tested whether there is a lag between the X-ray and UV light curves and found that the strongest correlation between the X-ray and UV light curves in run A is still at $<0.08$~days. On the other hand, we found a very weak correlation for run B, $r=0.048$ and $p=0.017$, occurs for a lag of 27~days, which is roughly the thermal timescale near the surface of the disk. However, a lag of around 30 days for reprocessing has not been observed in disk continuum reverberation mapping campaigns. 

Overall, our simulation shows that the low absorption opacity for the hard X-ray irradiation makes it unlikely that hard X-rays are the main driver of variability in emitted UV light curves. Several AGN reverberation mapping campaigns have found no correlation \citep[e.g.,]{Morales:2019,Buisson:2018} or only weak correlations \citep[e.g.,][]{Schimoia:2015,Edelson2019,Cackett:2023} between X-ray and UV-optical light curves, which agrees with our results here. However, it is possible that this observed lack of correlation in these papers is simply due to absorption of intervening material \citep[e.g.,][]{Kara:2021}.

\section{Summary and Future Work}
\label{sec:discuss}

Using the multi-frequency radiation MHD code from \cite{Jiang2022} allows us to simulate the reprocessing of X-ray light curves into UV light curves from first principles. This reprocessing is a key component of reverberation mapping, one of the primary tools used to study AGN disks. The traditional model for reverberation mapping assumes that X-rays are the main driver of variability in UV and optical light curves emitted by the AGN disk. However, the results from our simulations do not agree with this assumption. Instead we find that there is significant intrinsic UV variability locally generated by MRI and convection. This variability can be described by a DRW, with $\tau_{\rm damp}\approx30$~days on timescales greater than five days, but shows less power than the DRW at higher frequency, in agreement with AGN light curves observed at high cadence by Kepler \citep{Smith2018}. 

The X-rays we inject into our simulations have almost no impact on the PSDs of the reprocessed UV light curves in our two-frequency runs, because the short timescale variability of the X-ray irradiation gets smoothed when it is scattered in the outer regions of the disk. As a result, the X-ray and UV light curves from our two-frequency runs will not be well-correlated if there is not sufficient X-ray variability on timescales greater than a day. Even with longer timescale variability, we only find a correlation of $r=0.52$ between the X-ray and UV light curves, again because of the low absorption opacity for X-rays in our simulations. This correlation agrees with observational results that suggest that X-ray and UV light curves are not as well-correlated as we might expect if X-rays are the main driver of UV and optical variability \citep[e.g.,][]{Schimoia:2015,Edelson:2019,Cackett:2023}.

There are many radiation MHD simulations of AGN disks, but most are one-frequency simulations and none focus on the impact of X-ray irradiation on the UV-optical region of AGN disks and the light curves emitted by the disk. On the other hand, there are numerous sophisticated (semi-)analytical calculations for disk reprocessing of X-ray irradiation  \citep[e.g.,][]{Sun:2020,Kammoun:2021b,Salvesen:2022,Panagiotou:2022}. However, they all make assumptions about the vertical profiles of AGN disk parameters and do not evolve the gas turbulence in three dimensions nor the impact that radiation has on the gas and vice versa as a function of time. Doing so in the simulations presented here, provides detailed information on the nature of light curve variability produced by disk turbulence, and the ability of X-ray light curves with different PSDs to transfer energy and momentum to the AGN disk, affecting the variability of its emission.

Increasing our understanding of the physical processes that influence reverberation mapping is crucial because there is growing evidence that the standard \cite{Shakura1973} thin disk model does not accurately model an AGN disk. For example, radiation MHD simulations show that pressure support from magnetic fields and radiation can lead to puffed up disks \citep[e.g.,][]{Jiang2016,Jiang:2019,JiangBlaes2020,Hopkins:2023}, while recent observations provide additional evidence for thicker disks than predicted by the standard thin disk model \citep{Yao:2022,Secunda:2023}. In addition, both micro-lensing and reverberation mapping campaigns suggest that the radial extent of the disk is larger than predicted by the standard disk model \citep[e.g.,][]{Morgan:2010,Jiang:2017,Guo:2022}. 

Improved intuition about the physical processes involved in reverberation mapping is essential to improving our models for AGN disks. Future work should look at whether lower energy soft X-ray or far-UV irradiation, which will have larger absorption opacities compared to hard X-ray irradiation, are more likely to be absorbed and reprocessed by the disk producing stronger correlations between injected and reprocessed light curves. Future work should also look at a wider range of parameters for AGN disks in order to determine how the physics of reverberation mapping changes as a function of parameters such as SMBH mass and accretion rate. Finally, expanding these shearing box simulations to global simulations is necessary to understand how the fluctuations producing this intrinsic variability are propagated throughout the disk affecting the reverberation mapping signal. These simulations will allow us to make our reverberation mapping techniques more accurate and mine more information out of AGN light curves about the structure and internal physics of AGN disks.

\begin{acknowledgements}
    A.S. is supported by a fellowship from the NSF Graduate Research Fellowship Program under Grant No. DGE-1656466. JEG is supported in part by NSF grants AST1007052 and AST1007094. The Center for Computational Astrophysics at the Flatiron Institute is supported by the Simons Foundation.
\end{acknowledgements}

\bibliography{agn_lag.bib}

\begin{thebibliography}{}
\expandafter\ifx\csname natexlab\endcsname\relax\def\natexlab#1{#1}\fi
\providecommand{\url}[1]{\href{#1}{#1}}
\providecommand{\dodoi}[1]{doi:~\href{http://doi.org/#1}{\nolinkurl{#1}}}
\providecommand{\doeprint}[1]{\href{http://ascl.net/#1}{\nolinkurl{http://ascl.net/#1}}}
\providecommand{\doarXiv}[1]{\href{https://arxiv.org/abs/#1}{\nolinkurl{https://arxiv.org/abs/#1}}}

\bibitem[{{Ar{\'e}valo} {et~al.}(2009){Ar{\'e}valo}, {Uttley}, {Lira},
  {Breedt}, {McHardy}, \& {Churazov}}]{Arevalo:2009}
{Ar{\'e}valo}, P., {Uttley}, P., {Lira}, P., {et~al.} 2009, \mnras, 397, 2004,
  \dodoi{10.1111/j.1365-2966.2009.15110.x}

\bibitem[{{Bai} \& {Stone}(2013)}]{Bai:2013}
{Bai}, X.-N., \& {Stone}, J.~M. 2013, \apj, 769, 76,
  \dodoi{10.1088/0004-637X/769/1/76}

\bibitem[{{Balbus} \& {Hawley}(1991)}]{BalbusHawley1991}
{Balbus}, S.~A., \& {Hawley}, J.~F. 1991, \apj, 376, 214,
  \dodoi{10.1086/170270}

\bibitem[{Bentz \& Katz(2015)}]{Bentz_2015}
Bentz, M.~C., \& Katz, S. 2015, Publications of the Astronomical Society of the
  Pacific, 127, 67, \dodoi{10.1086/679601}

\bibitem[{{Blandford} \& {McKee}(1982)}]{Blandford1982}
{Blandford}, R.~D., \& {McKee}, C.~F. 1982, \apj, 255, 419,
  \dodoi{10.1086/159843}

\bibitem[{{Buisson} {et~al.}(2018){Buisson}, {Lohfink}, {Alston}, {Cackett},
  {Chiang}, {Dauser}, {De Marco}, {Fabian}, {Gallo}, {Garc{\'\i}a}, {Jiang},
  {Kara}, {Middleton}, {Miniutti}, {Parker}, {Pinto}, {Uttley}, {Walton}, \&
  {Wilkins}}]{Buisson:2018}
{Buisson}, D.~J.~K., {Lohfink}, A.~M., {Alston}, W.~N., {et~al.} 2018, \mnras,
  475, 2306, \dodoi{10.1093/mnras/sty008}

\bibitem[{{Burke} {et~al.}(2021){Burke}, {Shen}, {Blaes}, {Gammie}, {Horne},
  {Jiang}, {Liu}, {McHardy}, {Morgan}, {Scaringi}, \& {Yang}}]{Burke2021}
{Burke}, C.~J., {Shen}, Y., {Blaes}, O., {et~al.} 2021, Science, 373, 789,
  \dodoi{10.1126/science.abg9933}

\bibitem[{{Cackett} {et~al.}(2021){Cackett}, {Bentz}, \& {Kara}}]{Cackett:2021}
{Cackett}, E.~M., {Bentz}, M.~C., \& {Kara}, E. 2021, iScience, 24, 102557,
  \dodoi{10.1016/j.isci.2021.102557}

\bibitem[{{Cackett} {et~al.}(2018){Cackett}, {Chiang}, {McHardy}, {Edelson},
  {Goad}, {Horne}, \& {Korista}}]{Cackett:2018}
{Cackett}, E.~M., {Chiang}, C.-Y., {McHardy}, I., {et~al.} 2018, \apj, 857, 53,
  \dodoi{10.3847/1538-4357/aab4f7}

\bibitem[{{Cackett} {et~al.}(2007){Cackett}, {Horne}, \&
  {Winkler}}]{Cackett:2007}
{Cackett}, E.~M., {Horne}, K., \& {Winkler}, H. 2007, \mnras, 380, 669,
  \dodoi{10.1111/j.1365-2966.2007.12098.x}

\bibitem[{{Cackett} {et~al.}(2023){Cackett}, {Gelbord}, {Barth}, {De Rosa},
  {Edelson}, {Goad}, {Homayouni}, {Horne}, {Kara}, {Kriss}, {Korista}, {Landt},
  {Plesha}, {Arav}, {Bentz}, {Boizelle}, {Dalla Bonta}, {Dehghanian}, {Donnan},
  {Du}, {Ferland}, {Fian}, {Filippenko}, {Gonzalez Buitrago}, {Grier}, {Hall},
  {Hu}, {Ilic}, {Kaastra}, {Kaspi}, {Kochanek}, {Kovacevic}, {Kynoch}, {Li},
  {McLane}, {Mehdipour}, {Miller}, {Montano}, {Netzer}, {Panagiotou},
  {Partington}, {Popovic}, {Proga}, {Rogantini}, {Sanmartim}, {Siebert},
  {Storchi-Bergmann}, {Vestergaard}, {Wang}, {Waters}, \&
  {Zaidouni}}]{Cackett:2023}
{Cackett}, E.~M., {Gelbord}, J., {Barth}, A.~J., {et~al.} 2023, arXiv e-prints,
  arXiv:2306.17663, \dodoi{10.48550/arXiv.2306.17663}

\bibitem[{{De Rosa} {et~al.}(2015){De Rosa}, {Peterson}, {Ely}, {Kriss},
  {Crenshaw}, {Horne}, {Korista}, {Netzer}, {Pogge}, {Ar{\'e}valo}, {Barth},
  {Bentz}, {Brandt}, {Breeveld}, {Brewer}, {Dalla Bont{\`a}}, {De
  Lorenzo-C{\'a}ceres}, {Denney}, {Dietrich}, {Edelson}, {Evans}, {Fausnaugh},
  {Gehrels}, {Gelbord}, {Goad}, {Grier}, {Grupe}, {Hall}, {Kaastra}, {Kelly},
  {Kennea}, {Kochanek}, {Lira}, {Mathur}, {McHardy}, {Nousek}, {Pancoast},
  {Papadakis}, {Pei}, {Schimoia}, {Siegel}, {Starkey}, {Treu}, {Uttley},
  {Vaughan}, {Vestergaard}, {Villforth}, {Yan}, {Young}, \& {Zu}}]{derosa2015}
{De Rosa}, G., {Peterson}, B.~M., {Ely}, J., {et~al.} 2015, \apj, 806, 128,
  \dodoi{10.1088/0004-637X/806/1/128}

\bibitem[{{Edelson} {et~al.}(2015){Edelson}, {Gelbord}, {Horne}, {McHardy},
  {Peterson}, {Ar{\'e}valo}, {Breeveld}, {De Rosa}, {Evans}, {Goad}, {Kriss},
  {Brandt}, {Gehrels}, {Grupe}, {Kennea}, {Kochanek}, {Nousek}, {Papadakis},
  {Siegel}, {Starkey}, {Uttley}, {Vaughan}, {Young}, {Barth}, {Bentz},
  {Brewer}, {Crenshaw}, {Dalla Bont{\`a}}, {De Lorenzo-C{\'a}ceres}, {Denney},
  {Dietrich}, {Ely}, {Fausnaugh}, {Grier}, {Hall}, {Kaastra}, {Kelly},
  {Korista}, {Lira}, {Mathur}, {Netzer}, {Pancoast}, {Pei}, {Pogge},
  {Schimoia}, {Treu}, {Vestergaard}, {Villforth}, {Yan}, \& {Zu}}]{Edelson2015}
{Edelson}, R., {Gelbord}, J.~M., {Horne}, K., {et~al.} 2015, \apj, 806, 129,
  \dodoi{10.1088/0004-637X/806/1/129}

\bibitem[{{Edelson} {et~al.}(2017){Edelson}, {Gelbord}, {Cackett}, {Connolly},
  {Done}, {Fausnaugh}, {Gardner}, {Gehrels}, {Goad}, {Horne}, {McHardy},
  {Peterson}, {Vaughan}, {Vestergaard}, {Breeveld}, {Barth}, {Bentz},
  {Bottorff}, {Brandt}, {Crawford}, {Dalla Bont{\`a}}, {Emmanoulopoulos},
  {Evans}, {Figuera Jaimes}, {Filippenko}, {Ferland}, {Grupe}, {Joner},
  {Kennea}, {Korista}, {Krimm}, {Kriss}, {Leonard}, {Mathur}, {Netzer},
  {Nousek}, {Page}, {Romero-Colmenero}, {Siegel}, {Starkey}, {Treu}, {Vogler},
  {Winkler}, \& {Zheng}}]{Edelson2017}
{Edelson}, R., {Gelbord}, J., {Cackett}, E., {et~al.} 2017, \apj, 840, 41,
  \dodoi{10.3847/1538-4357/aa6890}

\bibitem[{{Edelson} {et~al.}(2019{\natexlab{a}}){Edelson}, {Gelbord},
  {Cackett}, {Peterson}, {Horne}, {Barth}, {Starkey}, {Bentz}, {Brandt},
  {Goad}, {Joner}, {Korista}, {Netzer}, {Page}, {Uttley}, {Vaughan},
  {Breeveld}, {Cenko}, {Done}, {Evans}, {Fausnaugh}, {Ferland},
  {Gonzalez-Buitrago}, {Gropp}, {Grupe}, {Kaastra}, {Kennea}, {Kriss},
  {Mathur}, {Mehdipour}, {Mudd}, {Nousek}, {Schmidt}, {Vestergaard}, \&
  {Villforth}}]{Edelson2019}
---. 2019{\natexlab{a}}, \apj, 870, 123, \dodoi{10.3847/1538-4357/aaf3b4}

\bibitem[{{Edelson} {et~al.}(2019{\natexlab{b}}){Edelson}, {Gelbord},
  {Cackett}, {Peterson}, {Horne}, {Barth}, {Starkey}, {Bentz}, {Brandt},
  {Goad}, {Joner}, {Korista}, {Netzer}, {Page}, {Uttley}, {Vaughan},
  {Breeveld}, {Cenko}, {Done}, {Evans}, {Fausnaugh}, {Ferland},
  {Gonzalez-Buitrago}, {Gropp}, {Grupe}, {Kaastra}, {Kennea}, {Kriss},
  {Mathur}, {Mehdipour}, {Mudd}, {Nousek}, {Schmidt}, {Vestergaard}, \&
  {Villforth}}]{Edelson:2019}
---. 2019{\natexlab{b}}, \apj, 870, 123, \dodoi{10.3847/1538-4357/aaf3b4}

\bibitem[{{Fausnaugh} {et~al.}(2016){Fausnaugh}, {Denney}, {Barth}, {Bentz},
  {Bottorff}, {Carini}, {Croxall}, {De Rosa}, {Goad}, {Horne}, {Joner},
  {Kaspi}, {Kim}, {Klimanov}, {Kochanek}, {Leonard}, {Netzer}, {Peterson},
  {Schn{\"u}lle}, {Sergeev}, {Vestergaard}, {Zheng}, {Zu}, {Anderson},
  {Ar{\'e}valo}, {Bazhaw}, {Borman}, {Boroson}, {Brandt}, {Breeveld}, {Brewer},
  {Cackett}, {Crenshaw}, {Dalla Bont{\`a}}, {De Lorenzo-C{\'a}ceres},
  {Dietrich}, {Edelson}, {Efimova}, {Ely}, {Evans}, {Filippenko}, {Flatland},
  {Gehrels}, {Geier}, {Gelbord}, {Gonzalez}, {Gorjian}, {Grier}, {Grupe},
  {Hall}, {Hicks}, {Horenstein}, {Hutchison}, {Im}, {Jensen}, {Jones},
  {Kaastra}, {Kelly}, {Kennea}, {Kim}, {Korista}, {Kriss}, {Lee}, {Lira},
  {MacInnis}, {Manne-Nicholas}, {Mathur}, {McHardy}, {Montouri}, {Musso},
  {Nazarov}, {Norris}, {Nousek}, {Okhmat}, {Pancoast}, {Papadakis}, {Parks},
  {Pei}, {Pogge}, {Pott}, {Rafter}, {Rix}, {Saylor}, {Schimoia}, {Siegel},
  {Spencer}, {Starkey}, {Sung}, {Teems}, {Treu}, {Turner}, {Uttley},
  {Villforth}, {Weiss}, {Woo}, {Yan}, \& {Young}}]{Fausnaugh2016}
{Fausnaugh}, M.~M., {Denney}, K.~D., {Barth}, A.~J., {et~al.} 2016, \apj, 821,
  56, \dodoi{10.3847/0004-637X/821/1/56}

\bibitem[{{Garc{\'\i}a} {et~al.}(2013){Garc{\'\i}a}, {Dauser}, {Reynolds},
  {Kallman}, {McClintock}, {Wilms}, \& {Eikmann}}]{Garcia:2013}
{Garc{\'\i}a}, J., {Dauser}, T., {Reynolds}, C.~S., {et~al.} 2013, \apj, 768,
  146, \dodoi{10.1088/0004-637X/768/2/146}

\bibitem[{{Gardner} \& {Done}(2017)}]{Gardner:2017}
{Gardner}, E., \& {Done}, C. 2017, \mnras, 470, 3591,
  \dodoi{10.1093/mnras/stx946}

\bibitem[{Grier {et~al.}(2017)Grier, Trump, Shen, Horne, Kinemuchi, McGreer,
  Starkey, Brandt, Hall, Kochanek, Chen, Denney, Greene, Ho, Homayouni, Li,
  Pei, Peterson, Petitjean, Schneider, Sun, AlSayyad, Bizyaev, Brinkmann,
  Brownstein, Bundy, Dawson, Eftekharzadeh, Fernandez-Trincado, Gao,
  Hutchinson, Jia, Jiang, Oravetz, Pan, Paris, Ponder, Peters, Rogerson,
  Simmons, Smith, , \& Wang}]{Grier:2017}
Grier, C.~J., Trump, J.~R., Shen, Y., {et~al.} 2017, The Astrophysical Journal,
  851, 21, \dodoi{10.3847/1538-4357/aa98dc}

\bibitem[{{Guo} {et~al.}(2022){Guo}, {Barth}, \& {Wang}}]{Guo:2022}
{Guo}, H., {Barth}, A.~J., \& {Wang}, S. 2022, arXiv e-prints,
  arXiv:2207.06432.
\newblock \doarXiv{2207.06432}

\bibitem[{{Hagen} \& {Done}(2023)}]{Hagen2023}
{Hagen}, S., \& {Done}, C. 2023, \mnras, 521, 251,
  \dodoi{10.1093/mnras/stad504}

\bibitem[{{Hall} {et~al.}(2018){Hall}, {Sarrouh}, \& {Horne}}]{Hall:2018}
{Hall}, P.~B., {Sarrouh}, G.~T., \& {Horne}, K. 2018, \apj, 854, 93,
  \dodoi{10.3847/1538-4357/aaa768}

\bibitem[{{Homayouni} {et~al.}(2022){Homayouni}, {Sturm}, {Trump}, {Horne},
  {Grier}, {Shen}, {Brandt}, {Alvarez}, {Hall}, {Ho}, {Li}, {Sun}, \&
  {Schneider}}]{Homayouni:2022}
{Homayouni}, Y., {Sturm}, M.~R., {Trump}, J.~R., {et~al.} 2022, \apj, 926, 225,
  \dodoi{10.3847/1538-4357/ac478b}

\bibitem[{{Hopkins} {et~al.}(2023){Hopkins}, {Squire}, {Su}, {Steinwandel},
  {Kremer}, {Shi}, {Grudic}, {Wellons}, {Faucher-Giguere}, {Angles-Alcazar},
  {Murray}, \& {Quataert}}]{Hopkins:2023}
{Hopkins}, P.~F., {Squire}, J., {Su}, K.-Y., {et~al.} 2023, arXiv e-prints,
  arXiv:2310.04506, \dodoi{10.48550/arXiv.2310.04506}

\bibitem[{{Iglesias} \& {Rogers}(1996)}]{IglesiasRogers1996}
{Iglesias}, C.~A., \& {Rogers}, F.~J. 1996, \apj, 464, 943,
  \dodoi{10.1086/177381}

\bibitem[{{Jiang}(2021)}]{Jiang2021}
{Jiang}, Y.-F. 2021, \apjs, 253, 49, \dodoi{10.3847/1538-4365/abe303}

\bibitem[{{Jiang}(2022)}]{Jiang2022}
---. 2022, \apjs, 263, 4, \dodoi{10.3847/1538-4365/ac9231}

\bibitem[{{Jiang} \& {Blaes}(2020)}]{JiangBlaes2020}
{Jiang}, Y.-F., \& {Blaes}, O. 2020, \apj, 900, 25,
  \dodoi{10.3847/1538-4357/aba4b7}

\bibitem[{{Jiang} {et~al.}(2019){Jiang}, {Blaes}, {Stone}, \&
  {Davis}}]{Jiang:2019}
{Jiang}, Y.-F., {Blaes}, O., {Stone}, J.~M., \& {Davis}, S.~W. 2019, ApJ, 885,
  144, \dodoi{10.3847/1538-4357/ab4a00}

\bibitem[{{Jiang} {et~al.}(2016){Jiang}, {Davis}, \& {Stone}}]{Jiang2016}
{Jiang}, Y.-F., {Davis}, S.~W., \& {Stone}, J.~M. 2016, \apj, 827, 10,
  \dodoi{10.3847/0004-637X/827/1/10}

\bibitem[{{Jiang} {et~al.}(2013){Jiang}, {Stone}, \& {Davis}}]{Jiang2013}
{Jiang}, Y.-F., {Stone}, J.~M., \& {Davis}, S.~W. 2013, \apj, 778, 65,
  \dodoi{10.1088/0004-637X/778/1/65}

\bibitem[{{Jiang} {et~al.}(2017){Jiang}, {Green}, {Greene}, {Morganson},
  {Shen}, {Pancoast}, {MacLeod}, {Anderson}, {Brandt}, {Grier}, {Rix}, {Ruan},
  {Protopapas}, {Scott}, {Burgett}, {Hodapp}, {Huber}, {Kaiser}, {Kudritzki},
  {Magnier}, {Metcalfe}, {Tonry}, {Wainscoat}, \& {Waters}}]{Jiang:2017}
{Jiang}, Y.-F., {Green}, P.~J., {Greene}, J.~E., {et~al.} 2017, \apj, 836, 186,
  \dodoi{10.3847/1538-4357/aa5b91}

\bibitem[{{Kammoun} {et~al.}(2021){Kammoun}, {Papadakis}, \&
  {Dov{\v{c}}iak}}]{Kammoun:2021b}
{Kammoun}, E.~S., {Papadakis}, I.~E., \& {Dov{\v{c}}iak}, M. 2021, \mnras, 503,
  4163, \dodoi{10.1093/mnras/stab725}

\bibitem[{{Kara} {et~al.}(2021){Kara}, {Mehdipour}, {Kriss}, {Cackett}, {Arav},
  {Barth}, {Byun}, {Brotherton}, {De Rosa}, {Gelbord}, {Hern{\'a}ndez
  Santisteban}, {Hu}, {Kaastra}, {Landt}, {Li}, {Miller}, {Montano},
  {Partington}, {Aceituno}, {Bai}, {Bao}, {Bentz}, {Brink}, {Chelouche},
  {Chen}, {Colmenero}, {Dalla Bont{\`a}}, {Dehghanian}, {Du}, {Edelson},
  {Ferland}, {Ferrarese}, {Fian}, {Filippenko}, {Fischer}, {Goad},
  {Gonz{\'a}lez Buitrago}, {Gorjian}, {Grier}, {Guo}, {Hall}, {Ho},
  {Homayouni}, {Horne}, {Ili{\'c}}, {Jiang}, {Joner}, {Kaspi}, {Kochanek},
  {Korista}, {Kynoch}, {Li}, {Liu}, {McHardy}, {McLane}, {Mitchell}, {Netzer},
  {Olson}, {Pogge}, {Popovi{\'c}}, {Proga}, {Storchi-Bergmann}, {Strasburger},
  {Treu}, {Vestergaard}, {Wang}, {Ward}, {Waters}, {Williams}, {Yang}, {Yao},
  {Zastrocky}, {Zhai}, \& {Zu}}]{Kara:2021}
{Kara}, E., {Mehdipour}, M., {Kriss}, G.~A., {et~al.} 2021, \apj, 922, 151,
  \dodoi{10.3847/1538-4357/ac2159}

\bibitem[{{Kara} {et~al.}(2023){Kara}, {Barth}, {Cackett}, {Gelbord},
  {Montano}, {Li}, {Santana}, {Horne}, {Alston}, {Buisson}, {Chelouche}, {Du},
  {Fabian}, {Fian}, {Gallo}, {Goad}, {Grupe}, {Gonz{\'a}lez Buitrago},
  {Hern{\'a}ndez Santisteban}, {Kaspi}, {Hu}, {Komossa}, {Kriss}, {Lewin},
  {Lewis}, {Loewenstein}, {Lohfink}, {Masterson}, {McHardy}, {Mehdipour},
  {Miller}, {Panagiotou}, {Parker}, {Pinto}, {Remillard}, {Reynolds},
  {Rogantini}, {Wang}, {Wang}, \& {Wilkins}}]{Kara:2023}
{Kara}, E., {Barth}, A.~J., {Cackett}, E.~M., {et~al.} 2023, ApJ, 947, 62,
  \dodoi{10.3847/1538-4357/acbcd3}

\bibitem[{Kaspi {et~al.}(2000)Kaspi, Smith, Netzer, Maoz, Jannuzi, \&
  Giveon}]{Kaspi_2000}
Kaspi, S., Smith, P.~S., Netzer, H., {et~al.} 2000, The Astrophysical Journal,
  533, 631, \dodoi{10.1086/308704}

\bibitem[{{Kelly} {et~al.}(2009{\natexlab{a}}){Kelly}, {Bechtold}, \&
  {Siemiginowska}}]{Kelly2009}
{Kelly}, B.~C., {Bechtold}, J., \& {Siemiginowska}, A. 2009{\natexlab{a}},
  \apj, 698, 895, \dodoi{10.1088/0004-637X/698/1/895}

\bibitem[{{Kelly} {et~al.}(2009{\natexlab{b}}){Kelly}, {Bechtold}, \&
  {Siemiginowska}}]{Kelly:2009}
---. 2009{\natexlab{b}}, \apj, 698, 895, \dodoi{10.1088/0004-637X/698/1/895}

\bibitem[{{Koz{\l}owski} {et~al.}(2010){Koz{\l}owski}, {Kochanek}, {Udalski},
  {Wyrzykowski}, {Soszy{\'n}ski}, {Szyma{\'n}ski}, {Kubiak}, {Pietrzy{\'n}ski},
  {Szewczyk}, {Ulaczyk}, {Poleski}, \& {OGLE Collaboration}}]{Kozlowski2010}
{Koz{\l}owski}, S., {Kochanek}, C.~S., {Udalski}, A., {et~al.} 2010, \apj, 708,
  927, \dodoi{10.1088/0004-637X/708/2/927}

\bibitem[{{MacLeod} {et~al.}(2012){MacLeod}, {Ivezi{\'c}}, {Sesar}, {de Vries},
  {Kochanek}, {Kelly}, {Becker}, {Lupton}, {Hall}, {Richards}, {Anderson}, \&
  {Schneider}}]{MacLeod2012}
{MacLeod}, C.~L., {Ivezi{\'c}}, {\v{Z}}., {Sesar}, B., {et~al.} 2012, \apj,
  753, 106, \dodoi{10.1088/0004-637X/753/2/106}

\bibitem[{{Morales} {et~al.}(2019){Morales}, {Miller}, {Cackett}, {Reynolds},
  \& {Zoghbi}}]{Morales:2019}
{Morales}, A.~M., {Miller}, J.~M., {Cackett}, E.~M., {Reynolds}, M.~T., \&
  {Zoghbi}, A. 2019, \apj, 870, 54, \dodoi{10.3847/1538-4357/aaeff9}

\bibitem[{{Morgan} {et~al.}(2010){Morgan}, {Kochanek}, {Morgan}, \&
  {Falco}}]{Morgan:2010}
{Morgan}, C.~W., {Kochanek}, C.~S., {Morgan}, N.~D., \& {Falco}, E.~E. 2010,
  \apj, 712, 1129, \dodoi{10.1088/0004-637X/712/2/1129}

\bibitem[{{Panagiotou} {et~al.}(2022){Panagiotou}, {Papadakis}, {Kara},
  {Kammoun}, \& {Dov{\v{c}}iak}}]{Panagiotou:2022}
{Panagiotou}, C., {Papadakis}, I., {Kara}, E., {Kammoun}, E., \&
  {Dov{\v{c}}iak}, M. 2022, \apj, 935, 93, \dodoi{10.3847/1538-4357/ac7e4d}

\bibitem[{{Peterson}(2014)}]{Peterson:2014}
{Peterson}, B.~M. 2014, \ssr, 183, 253, \dodoi{10.1007/s11214-013-9987-4}

\bibitem[{{Peterson} {et~al.}(2004){Peterson}, {Ferrarese}, {Gilbert}, {Kaspi},
  {Malkan}, {Maoz}, {Merritt}, {Netzer}, {Onken}, {Pogge}, {Vestergaard}, \&
  {Wandel}}]{Peterson2004}
{Peterson}, B.~M., {Ferrarese}, L., {Gilbert}, K.~M., {et~al.} 2004, \apj, 613,
  682, \dodoi{10.1086/423269}

\bibitem[{{Salvesen}(2022)}]{Salvesen:2022}
{Salvesen}, G. 2022, \apjl, 940, L22, \dodoi{10.3847/2041-8213/ac9cdd}

\bibitem[{{Schimoia} {et~al.}(2015){Schimoia}, {Storchi-Bergmann}, {Grupe},
  {Eracleous}, {Peterson}, {Baldwin}, {Nemmen}, \& {Winge}}]{Schimoia:2015}
{Schimoia}, J.~S., {Storchi-Bergmann}, T., {Grupe}, D., {et~al.} 2015, \apj,
  800, 63, \dodoi{10.1088/0004-637X/800/1/63}

\bibitem[{{Secunda} {et~al.}(2023){Secunda}, {Greene}, {Jiang}, {Yao}, \&
  {Zoghbi}}]{Secunda:2023}
{Secunda}, A., {Greene}, J.~E., {Jiang}, Y.-F., {Yao}, P.~Z., \& {Zoghbi}, A.
  2023, arXiv e-prints, arXiv:2306.05455, \dodoi{10.48550/arXiv.2306.05455}

\bibitem[{{Sergeev} {et~al.}(2005){Sergeev}, {Doroshenko}, {Golubinskiy},
  {Merkulova}, \& {Sergeeva}}]{Sergeev:2005}
{Sergeev}, S.~G., {Doroshenko}, V.~T., {Golubinskiy}, Y.~V., {Merkulova},
  N.~I., \& {Sergeeva}, E.~A. 2005, \apj, 622, 129, \dodoi{10.1086/427820}

\bibitem[{{Shakura} \& {Sunyaev}(1973)}]{Shakura1973}
{Shakura}, N.~I., \& {Sunyaev}, R.~A. 1973, \aap, 24, 337

\bibitem[{{Smith} {et~al.}(2018){Smith}, {Mushotzky}, {Boyd}, {Malkan},
  {Howell}, \& {Gelino}}]{Smith2018}
{Smith}, K.~L., {Mushotzky}, R.~F., {Boyd}, P.~T., {et~al.} 2018, \apj, 857,
  141, \dodoi{10.3847/1538-4357/aab88d}

\bibitem[{{Starkey} {et~al.}(2017){Starkey}, {Horne}, {Fausnaugh}, {Peterson},
  {Bentz}, {Kochanek}, {Denney}, {Edelson}, {Goad}, {De Rosa}, {Anderson},
  {Ar{\'e}valo}, {Barth}, {Bazhaw}, {Borman}, {Boroson}, {Bottorff}, {Brandt},
  {Breeveld}, {Cackett}, {Carini}, {Croxall}, {Crenshaw}, {Dalla Bont{\`a}},
  {De Lorenzo-C{\'a}ceres}, {Dietrich}, {Efimova}, {Ely}, {Evans},
  {Filippenko}, {Flatland}, {Gehrels}, {Geier}, {Gelbord}, {Gonzalez},
  {Gorjian}, {Grier}, {Grupe}, {Hall}, {Hicks}, {Horenstein}, {Hutchison},
  {Im}, {Jensen}, {Joner}, {Jones}, {Kaastra}, {Kaspi}, {Kelly}, {Kennea},
  {Kim}, {Kim}, {Klimanov}, {Korista}, {Kriss}, {Lee}, {Leonard}, {Lira},
  {MacInnis}, {Manne-Nicholas}, {Mathur}, {McHardy}, {Montouri}, {Musso},
  {Nazarov}, {Norris}, {Nousek}, {Okhmat}, {Pancoast}, {Parks}, {Pei}, {Pogge},
  {Pott}, {Rafter}, {Rix}, {Saylor}, {Schimoia}, {Schn{\"u}lle}, {Sergeev},
  {Siegel}, {Spencer}, {Sung}, {Teems}, {Turner}, {Uttley}, {Vestergaard},
  {Villforth}, {Weiss}, {Woo}, {Yan}, {Young}, {Zheng}, \& {Zu}}]{Starkey2017}
{Starkey}, D., {Horne}, K., {Fausnaugh}, M.~M., {et~al.} 2017, \apj, 835, 65,
  \dodoi{10.3847/1538-4357/835/1/65}

\bibitem[{{Stone} {et~al.}(2022){Stone}, {Shen}, {Burke}, {Chen}, {Yang},
  {Liu}, {Gruendl}, {Adam{\'o}w}, {Andrade-Oliveira}, {Annis}, {Bacon},
  {Bertin}, {Bocquet}, {Brooks}, {Burke}, {Carnero Rosell}, {Carrasco Kind},
  {Carretero}, {da Costa}, {Pereira}, {De Vicente}, {Desai}, {Diehl}, {Doel},
  {Ferrero}, {Friedel}, {Frieman}, {Garc{\'\i}a-Bellido}, {Gaztanaga}, {Gruen},
  {Gutierrez}, {Hinton}, {Hollowood}, {Honscheid}, {James}, {Kuehn},
  {Kuropatkin}, {Lidman}, {Maia}, {Menanteau}, {Miquel}, {Morgan},
  {Paz-Chinch{\'o}n}, {Pieres}, {Plazas Malag{\'o}n}, {Rodriguez-Monroy},
  {Sanchez}, {Scarpine}, {Serrano}, {Sevilla-Noarbe}, {Smith}, {Suchyta},
  {Swanson}, {Tarl{\'e}}, {To}, \& {DES Collaboration}}]{Stone:2022}
{Stone}, Z., {Shen}, Y., {Burke}, C.~J., {et~al.} 2022, \mnras, 514, 164,
  \dodoi{10.1093/mnras/stac1259}

\bibitem[{{Sun} {et~al.}(2020){Sun}, {Xue}, {Guo}, {Wang}, {Brandt}, {Trump},
  {He}, {Liu}, {Wu}, \& {Li}}]{Sun:2020}
{Sun}, M., {Xue}, Y., {Guo}, H., {et~al.} 2020, \apj, 902, 7,
  \dodoi{10.3847/1538-4357/abb1c4}

\bibitem[{{Tachibana} {et~al.}(2020){Tachibana}, {Graham}, {Kawai},
  {Djorgovski}, {Drake}, {Mahabal}, \& {Stern}}]{Tachibana:2020}
{Tachibana}, Y., {Graham}, M.~J., {Kawai}, N., {et~al.} 2020, \apj, 903, 54,
  \dodoi{10.3847/1538-4357/abb9a9}

\bibitem[{{Uttley} {et~al.}(2003){Uttley}, {Edelson}, {McHardy}, {Peterson}, \&
  {Markowitz}}]{Uttley:2003}
{Uttley}, P., {Edelson}, R., {McHardy}, I.~M., {Peterson}, B.~M., \&
  {Markowitz}, A. 2003, \apjl, 584, L53, \dodoi{10.1086/373887}

\bibitem[{{Yao} {et~al.}(2022){Yao}, {Secunda}, {Jiang}, {Greene}, \&
  {Villar}}]{Yao:2022}
{Yao}, P.~Z., {Secunda}, A., {Jiang}, Y.-F., {Greene}, J.~E., \& {Villar}, A.
  2022, arXiv e-prints, arXiv:2210.13489.
\newblock \doarXiv{2210.13489}

\bibitem[{{Yu} {et~al.}(2022){Yu}, {Richards}, {Vogeley}, {Moreno}, \&
  {Graham}}]{Yu:2022}
{Yu}, W., {Richards}, G.~T., {Vogeley}, M.~S., {Moreno}, J., \& {Graham}, M.~J.
  2022, \apj, 936, 132, \dodoi{10.3847/1538-4357/ac8351}

\bibitem[{{Zu} {et~al.}(2013){Zu}, {Kochanek}, {Koz{\l}owski}, \&
  {Udalski}}]{Zu2013}
{Zu}, Y., {Kochanek}, C.~S., {Koz{\l}owski}, S., \& {Udalski}, A. 2013, \apj,
  765, 106, \dodoi{10.1088/0004-637X/765/2/106}

\end{thebibliography}
\end{CJK*}
\end{document}